\catcode`@=11

\newskip\plaincentering \plaincentering=0pt plus 1000pt minus 1000pt
\def\@plainlign{\tabskip=0pt\everycr={}}
\def\eqalignno#1{\displ@y \tabskip\plaincentering
  \halign to\displaywidth{\hfil$\@lign\displaystyle{##}$\tabskip\z@skip
    &$\@lign\displaystyle{{}##}$\hfil\tabskip\plaincentering
    &\llap{$\@lign##$}\tabskip\z@skip\crcr
    #1\crcr}}
\def\leqalignno#1{\displ@y \tabskip\plaincentering
  \halign to\displaywidth{\hfil$\@lign\displaystyle{##}$\tabskip\z@skip
    &$\@lign\displaystyle{{}##}$\hfil\tabskip\plaincentering
    &\kern-\displaywidth\rlap{$\@lign##$}\tabskip\displaywidth\crcr
    #1\crcr}}
\def\plainLet@{\relax\iffalse{\fi\let\\=\cr\iffalse}\fi}
\def\plainvspace@{\def\vspace##1{\noalign{\vskip##1}}}

\def\intic@{\mathchoice{\hskip5\p@}{\hskip4\p@}{\hskip4\p@}{\hskip4\p@}}
\def\negintic@
 {\mathchoice{\hskip-5\p@}{\hskip-4\p@}{\hskip-4\p@}{\hskip-4\p@}}
\def\intkern@{\mathchoice{\!\!\!}{\!\!}{\!\!}{\!\!}}
\def\intdots@{\mathchoice{\cdots}{{\cdotp}\mkern1.5mu
    {\cdotp}\mkern1.5mu{\cdotp}}{{\cdotp}\mkern1mu{\cdotp}\mkern1mu
      {\cdotp}}{{\cdotp}\mkern1mu{\cdotp}\mkern1mu{\cdotp}}}
\newcount\intno@
\def\iint{\intno@=\tw@\futurelet\next\ints@}
\def\iiint{\intno@=\thr@@\futurelet\next\ints@}
\def\iiiint{\intno@=4 \futurelet\next\ints@}
\def\idotsint{\intno@=\z@\futurelet\next\ints@}
\def\ints@{\findlimits@\ints@@}
\newif\iflimtoken@
\newif\iflimits@
\def\findlimits@{\limtoken@false\limits@false\ifx\next\limits
 \limtoken@true\limits@true\else\ifx\next\nolimits\limtoken@true\limits@false
    \fi\fi}
\def\multintlimits@{\intop\ifnum\intno@=\z@\intdots@
  \else\intkern@\fi
    \ifnum\intno@>\tw@\intop\intkern@\fi
     \ifnum\intno@>\thr@@\intop\intkern@\fi\intop}
\def\multint@{\int\ifnum\intno@=\z@\intdots@\else\intkern@\fi
   \ifnum\intno@>\tw@\int\intkern@\fi
    \ifnum\intno@>\thr@@\int\intkern@\fi\int}
\def\ints@@{\iflimtoken@\def\ints@@@{\iflimits@
   \negintic@\mathop{\intic@\multintlimits@}\limits\else
    \multint@\nolimits\fi\eat@}\else
     \def\ints@@@{\multint@\nolimits}\fi\ints@@@}
\def\Sb{_\bgroup\vspace@
        \baselineskip=\fontdimen10 \scriptfont\tw@
        \advance\baselineskip by \fontdimen12 \scriptfont\tw@
        \lineskip=\thr@@\fontdimen8 \scriptfont\thr@@
        \lineskiplimit=\thr@@\fontdimen8 \scriptfont\thr@@
        \Let@\vbox\bgroup\halign\bgroup \hfil$\scriptstyle
            {##}$\hfil\cr}
\def\endSb{\crcr\egroup\egroup\egroup}
\def\Sp{^\bgroup\vspace@
        \baselineskip=\fontdimen10 \scriptfont\tw@
        \advance\baselineskip by \fontdimen12 \scriptfont\tw@
        \lineskip=\thr@@\fontdimen8 \scriptfont\thr@@
        \lineskiplimit=\thr@@\fontdimen8 \scriptfont\thr@@
        \Let@\vbox\bgroup\halign\bgroup \hfil$\scriptstyle
            {##}$\hfil\cr}
\def\endSp{\crcr\egroup\egroup\egroup}
\def\Let@{\relax\iffalse{\fi\let\\=\cr\iffalse}\fi}
\def\vspace@{\def\vspace##1{\noalign{\vskip##1 }}}
\def\aligned{\,\vcenter\bgroup\plainvspace@\plainLet@\openup\jot\m@th\ialign
  \bgroup \strut\hfil$\displaystyle{##}$&$\displaystyle{{}##}$\hfil\crcr}
\def\endaligned{\crcr\egroup\egroup}
\def\matrix{\,\vcenter\bgroup\plainLet@\plainvspace@
    \normalbaselines
  \m@th\ialign\bgroup\hfil$##$\hfil&&\quad\hfil$##$\hfil\crcr
    \mathstrut\crcr\noalign{\kern-\baselineskip}}
\def\endmatrix{\crcr\mathstrut\crcr\noalign{\kern-\baselineskip}\egroup
                \egroup\,}
\newtoks\hashtoks@
\hashtoks@={#}
\def\format{\crcr\egroup\iffalse{\fi\ifnum`}=0 \fi\format@}
\def\format@#1\\{\def\preamble@{#1}%
  \def\c{\hfil$\the\hashtoks@$\hfil}%
  \def\r{\hfil$\the\hashtoks@$}%
  \def\l{$\the\hashtoks@$\hfil}%
  \setbox\z@=\hbox{\xdef\Preamble@{\preamble@}}\ifnum`{=0 \fi\iffalse}\fi
   \ialign\bgroup\span\Preamble@\crcr}

\def\cases{\left\{\,\vcenter\bgroup\plainvspace@
     \normalbaselines\openup\jot\m@th
      \plainLet@\ialign\bgroup$\displaystyle{##}$\hfil&\quad$\displaystyle{{}##}$\hfil\crcr
      \mathstrut\crcr\noalign{\kern-\baselineskip}}

\newif\iftagsleft@
\tagsleft@true
\def\TagsOnRight{\global\tagsleft@false}
\def\tag#1$${\iftagsleft@\leqno\else\eqno\fi
 \hbox{\def\pagebreak{\global\postdisplaypenalty-\@M}%
 \def\nopagebreak{\global\postdisplaypenalty\@M}\rm(#1\unskip)}%
  $$\postdisplaypenalty\z@\ignorespaces}
\interdisplaylinepenalty=\@M
\def\plainallowdisplaybreak@{\def\allowdisplaybreak{\noalign{\allowbreak}}}
\def\plaindisplaybreak@{\def\displaybreak{\noalign{\break}}}
\def\align#1\endalign{\def\tag{&}\plainvspace@\plainallowdisplaybreak@\plaindisplaybreak@
  \iftagsleft@\plainlalign@#1\endalign\else
   \plainralign@#1\endalign\fi}
\def\plainralign@#1\endalign{\displ@y\plainLet@\tabskip\plaincentering\halign to\displaywidth
     {\hfil$\displaystyle{##}$\tabskip=\z@&$\displaystyle{{}##}$\hfil
       \tabskip=\plaincentering&\llap{\hbox{\rm(##\unskip)}}\tabskip\z@\crcr
             #1\crcr}}
\def\plainlalign@
 #1\endalign{\displ@y\plainLet@\tabskip\plaincentering\halign to \displaywidth
   {\hfil$\displaystyle{##}$\tabskip=\z@&$\displaystyle{{}##}$\hfil
   \tabskip=\plaincentering&\kern-\displaywidth
        \rlap{\hbox{\rm(##\unskip)}}\tabskip=\displaywidth\crcr
               #1\crcr}}

\def\re@#1{\par\hangindent\parindent\indent\llap{#1\enspace}\ignorespaces}
\def\qfootnote#1{\edef\@sf{\spacefactor\the\spacefactor}{}#1\@sf
      \insert\footins{\let\egroup=}\footnotesize 
      \interlinepenalty100 \let\par=\endgraf
        \leftskip=0pt \rightskip=0pt
        \splittopskip=10pt plus 1pt minus 1pt \floatingpenalty=20000
   \smallskip\re@{#1}\bgroup\strut\aftergroup{\strut\egroup}\let\next}
\catcode`\@=\active

\documentstyle[11pt]{article}
\begin{document}

\newcommand{\be}{\begin{equation}}
\newcommand{\ee}{\end{equation}}

\newcommand{\ba}{\begin{equation} \aligned}
\newcommand{\ea}{\endaligned \end{equation}}

\normalsize \baselineskip 18pt

\title{\bf  Ghost field realizations of the spinor $W_{2,s}$ strings based on
the linear $W_{1,2,s}$ algebras
           \thanks{Supported by the National Natural Science Foundation of China, grant No. 10275030.
            }}
\author{ Yuxiao Liu$\;\;$
         Lijie Zhang \thanks{E-mail address: zhanglj02@st.lzu.edu.cn} $\;\;$
         Jirong Ren $\;\;$
        \\
Institute of Theoretical Physics, Lanzhou University, Lanzhou
730000, P. R. China}
\date{}
\maketitle

\begin{center}
\begin{minipage}{120mm}
\vskip 0.5in

{\bf\noindent Abstract} \\

\noindent It has been shown that certain $W$ algebras can be
linearized by the inclusion of a spin-1 current. This Provides a
way of obtaining new realizations of the $W$ algebras. In this
paper, we investigate the new ghost field realizations of the
$W_{2,s}(s=3,4)$ algebras, making use of the fact that these two
algebras can be linearized. We then construct the nilpotent $BRST$
charges of the spinor non-critical $W_{2,s}$ strings with
these new realizations.\\

\vskip 15pt
\noindent {PACS: } 11.25.Sq, 11.10.-z, 11.25.Pm\\
\noindent {Keywords: } $W$ string, $BRST$ charge, Spinor realization\\

\end{minipage}
\end{center}

\newpage
{\bf\noindent 1. Introduction}

\indent  $W$ algebra has received considerable attention and
application since its discovery in the middle of the 1980's [1].
Much work has been carried out on the classification of it and the
study of the $W$ gravity and $W$ string theories. Furthermore, it
also appears in the quantum Hall effect, black holes, in lattice
models of statistical mechanics at criticality, and in other
physical models [2] and so on.\\
\indent In all applications of $W$ algebra, the investigation of
the $W$ string is more interesting and important. The $BRST$
method [3] has turned out to be the simplest way to study the
critical and non-critical $W$ string theories. The $BRST$ charge
of $W_{3}$ string was first constructed in [4], and the detailed
studies of it can be found in [4-6]. The scalar field realizations
of $W_{2,s}$ strings, which is a natural generalization of the
$W_{3}$ string, have been obtained for $s=4,5,6,7$ [6-8]. In Ref.
[9] we discovered the reason that the scalar $BRST$ charge is
difficult to be generalized to a general $W_{N}$ string. At the
same time, we found the methods to construct the spinor field
realizations of $W_{2,s}$ strings and $W_{N}$ strings [9,10].
Subsequently, we studied the exact spinor field realizations of
$W_{2,s}(s=3,4,5,6)$ strings and $W_{N}(N=4,5,6)$ strings [9-11].
Recently, the authors have constructed the nilpotent $BRST$
charges of spinor non-critical $W_{2,s}(s=3,4)$ strings by taking
into account the property of spinor field [12]. These results will
be of importance for constructing super $W$ strings, and they will
provide
the essential ingredients.\\
\indent However, all of these theories about the $W_{2,s}$ strings
mentioned above are based on the non-linear $W_{2,s}$ algebras.
Because of the intrinsic nonlinearity of the $W_{2,s}$ algebras,
their study is a more difficult task compared to linear algebras.
Fortunately, it has been shown that certain $W$ algebras can be
linearized by the inclusion of a spin-1 current. Once the linear
algebra is constructed, one could algorithmically reproduce the
structure of the corresponding nonlinear algebra. This provides a
way of obtaining new realizations of the $W_{2,s}$ algebras. Such
new realizations were constructed for the purpose of building the
corresponding scalar $W_{2,s}$ strings [13].\\
\indent Since up to now there is no work focused on the research
of ghost field realizations of the spinor $W_{2,s}$ strings based
on the linear $W_{1,2,s}$ algebras, we will construct the new
nilpotent $BRST$ charges of spinor non-critical $W_{2,s}$ strings
for the first time by using the linear bases of the $W_{1,2,s}$
algebras in this paper. To construct a non-critical $BRST$ charge
one must first solve the forms of matter currents $T$ and $W$
determined by the $OPEs$ of $TT$, $TW$ and $WW$. $T$ and $W$ here
are constructed by the linear bases $T_0, J_0, W_0$ of the
$W_{1,2,s}$ algebras, and these linear bases are constructed with
the ghost fields. Then direct substitution of these results into
$BRST$ charge leads to the grading spinor field realizations. Such
constructions are discussed for $s=3,4$ in detail. For the case of
critical realizations, the corresponding results can be obtained
naturally when the terms of matter currents vanish. All these
results will be of importance for embedding of the Virasoro string
into the $W_{2,s}$ strings.\\
\indent The present paper is organized as follows. In Section 2,
we construct the linear bases of the $W_{1,2,s}$ algebras and give
the ghost field realizations of the $W_{2,s}$ algebras. Then in
section 3, we build new $BRST$ charges of the spinor non-critical
$W_{2,3}$ and $W_{2,4}$ strings by using the grading
$BRST$ method.  And finally, the paper ends with a brief conclusion.\\

{\bf\noindent 2. The linear $W_{1,2,s}$ algebras and the new ghost
realizations of the $W_{2,s}$ algebras}

\indent The $W_{2,s}$ algebras can be linearized by the inclusion
of a spin-1 current [14]. The $OPEs$ of the spin-1 current with
the spin-2 current $T$ and spin-s current $W$ of the $W_{2,s}$
algebras are uniquely determined by requiring that the Jacobi
identities be satisfied. The bases $T$ and $W$ then can be
constructed by the linear bases of the $W_{1,2,s}$ algebras:
\ba T &=T_{0}\\
W &=W_{0}+W_{R}(J_{0},T_{0}) \ea

\noindent where the currents $T_{0},W_{0}$ and $J_{0}$ generate
the $W_{1,2,s}$ algebras. And the linearized $W_{1,2,3}$ and
$W_{1,2,4}$ algebras take the form: \ba T_{0}(z)T_{0}(\omega) \sim
& \frac{C/2}{(z - w)^4}+\frac{2T_{0}(\omega)}{(z-\omega)^2}+
\frac{\partial
T_{0}(\omega)}{(z-\omega)}, \\
 T_{0}(z)W_{0}(\omega) \sim &
\frac{s W_{0}(\omega)}{(z-\omega)^2}+ \frac{\partial
W_{0}(\omega)}{(z-\omega)},\\
T_{0}(z)J_{0}(\omega) \sim &  \frac{C_{1}}{(z -
w)^3}+\frac{J_{0}(\omega)}{(z-\omega)^2}+ \frac{\partial
J_{0}(\omega)}{(z-\omega)}, \\
J_{0}(z)J_{0}(\omega) \sim & \frac{-1}{(z-\omega)^2},\\
J_{0}(z)W_{0}(\omega) \sim & \frac{hW_{0}(\omega)}{(z-\omega)},\\
W_{0}(z)W_{0}(\omega) \sim & 0, \ea
 where $s=3$ and 4 respectively. The coefficients $C, C_{1}$ and
 $h$ are given by
 \ba
 C & =50+24t^{2}+\frac{24}{t^{2}},\;\;\;
 C_{1}=-\sqrt{6}(t+\frac{1}{t}),\;\;\;
 h=\sqrt{\frac{3}{2}}t, \;\;\; (s=3)\\
 C & =86+30t^{2}+\frac{60}{t^{2}},\;\;\;
 C_{1}=-3t-\frac{4}{t},\;\;\;\;\;\;\;\;\;
 h=t. \;\;\;\;\;\;\;\;\; (s=4)
 \ea

\indent One can realize the algebras given in Eq. (2) by the
two-spinor realizations in which the current $W_{0}$ is zero.
However, it was shown that $W_{0}$ does not have to be zero [14],
we can alternatively realize it in terms of the ghost field in
this paper. To obtain the new realizations for the linearized
$W_{1, 2, s}$ algebras, we use the bosonic ghost fields $(R,S)$
with spins $(s,1-s)$ and $(b_1,c_1)$ with spins $(k,1-k)$ to
construct the linear bases of them. Then the realizations for the
$W_{1,2,3}$ and $W_{1,2,4}$ algebras are given by \ba
T_0 =& T_{eff}+T_{g},\\
J_0 =& \rho R S+\lambda b_{1}c_{1},\\
W_{0}=&R,\\
\ea \noindent where \ba T_{g} =& s R \partial S+(s-1)\partial R
S+kb_{1} \partial c_{1}+(k-1) \partial b_{1} c_{1} \ea \noindent
and $T_{eff}$ is an arbitrary effective energy-momentum tensor
with central $C_{eff}$. By making use of the $OPEs$
$J_{0}(z)J_{0}(\omega)$ and $J_{0}(z)W_{0}(\omega)$ in (2), we can
solve the coefficients $\rho$ and $\lambda$. From the $OPE$
relation of $T_{0}$ and $J_{0}$ and Eq. (3), we determine the
value of $k$. Substituting this value into Eq. (5), we can
determine the central charge $C_g$ of $T_g$ with its $OPE$. Since
the central charge $C$ for $T_0$ is $C=C_g + C_{eff}$, the value
of $C_{eff}$ can be obtained. All the coefficients are listed as
follows:\\
\noindent (i)$\;\;$ s=3 \ba \rho=\sqrt{\frac{3}{2}}t, \;\;
\lambda=\pm\sqrt{1-\frac{3}{2} t^2},\;\;
k=\frac{1}{2}\mp\frac{\sqrt{6-9t^2}}{2t},\;\;C_g=46+\frac{18}{t^2},\;\;C_{eff}=4+\frac{6}{t^2}+24
t^2.\ea

\noindent (ii)$\;\;$ s=4 \ba \rho= -t,
\;\;\lambda=\pm\sqrt{1-t^2},\;\;
k=\frac{1}{2}\mp\frac{2\sqrt{1-t^2}}{t},\;\;C_g=97+\frac{48}{t^2},\;\;C_{eff}=-11+\frac{12}{t^2}+30
t^2. \ea

\indent Now let us turn our attention to the study of the ghost
field realizations of the $W_{2,s}$ algebras with linear bases of
the $W_{1,2,s}$ algebras. We begin by reviewing the structures of
the $W_{2,s}$ algebras in
conformal language. The $OPE$ $W(z)W(\omega)$ for $W_{2,3}$ algebra is given by [1]\\
 \ba W(z)W(\omega) \sim & \frac{C/3}{(z -
w)^6} +
\frac{2T}{(z-\omega)^4}+ \frac{\partial T}{(z-\omega)^3}\\
+& \frac{1}{(z-\omega)^2}(2 \Theta \Lambda +\frac{3}{10}\partial
^2 T )+\frac{1}{(z-\omega)}(\Theta
\partial \Lambda +\frac{1}{15}\partial ^3 T),
\ea
where\\
 \ba \Theta= \frac{16}{22+5C},\;\;\;\;\;\; \Lambda = T^2
- \frac{3}{10}
\partial ^2 T.
\ea

For the case $W_{2,4}$, the $OPE$ $W(z)W(\omega)$ take the form
[15]:
 \ba W(z)W(\omega) \sim &\{\frac{2 T}{(z - w)^6}+\frac{\partial
T}{(z-\omega)^5}+\frac{3}{10} \frac{\partial ^2
T}{(z-\omega)^4}\\
+ & \frac{1}{15}\frac{\partial ^3
T}{(z-\omega)^3}+\frac{1}{84}\frac{\partial ^4
T}{(z-\omega)^2}+\frac{1}{560}\frac{\partial ^5
T}{(z-\omega)}\}\\
+ & \sigma_{1} \{\frac{U}{(z-\omega)^4}+\frac{1}{2} \frac{\partial
U}{(z-\omega)^3}+ \frac{5}{36} \frac{\partial ^2
U}{(z-\omega)^2}+\frac{1}{36} \frac{\partial ^3
U}{(z-\omega)}\}\\
+ & \sigma_{2} \{ \frac{W}{(z-\omega)^4}+\frac{1}{2}
\frac{\partial W}{(z-\omega)^3}+ \frac{5}{36} \frac{\partial ^2
W}{(z-\omega)^2}+\frac{1}{36} \frac{\partial ^3 W}{(z-\omega)}\}\\
+& \sigma_{3} \{ \frac{G}{(z-\omega)^2}+\frac{1}{2}\frac{\partial
G}{(z-\omega)} \}+ \sigma_{4}\{ \frac{A}{(z-\omega)^2}+\frac{1}{2}
\frac{\partial A}{(z-\omega)}\}\\
+ & \sigma_{5}\{ \frac{B}{(z-\omega)^2}+\frac{1}{2} \frac{\partial
B}{(z-\omega)}\}+\frac{C/4}{(z-\omega)^8}, \ea \noindent where the
composites U (spin 4), and G, A and B (all spin 6 ), are defined
by \ba U & =(TT)-\frac{3}{10} \partial ^2 T,\;\;\;\;\;\;
G=(\partial ^2 T T)-\partial (\partial T T)+\frac{2}{9} \partial
^2
(TT)-\frac{1}{42}\partial ^4 T,\\
A & =(T U)-\frac{1}{6}\partial ^2 U, \;\;\;\;\;\; B=(T
W)-\frac{1}{6}\partial ^2 W ,\ea \noindent with normal ordering of
products of currents understood. The coefficients
$\sigma_{i}(i=1-5)$ are given by \ba
 \sigma_{1} & =\frac{42}{5C+22},\;\;\;\;\;\;
\sigma_{2}=\sqrt{\frac{54(C+24)(C^2 -172C
+196)}{(5C+22)(7C+68)(2C-1)}},\\
\sigma_{3} & =\frac{3(19C-524)}{10(7C+68)(2C-1)},\;\;\;\;\;\;
\sigma_{4}
=\frac{24(72C+13)}{(5C+22)(7C+68)(2C-1)},\\
\sigma_{5} & = \frac{28}{3(C+24)} \sigma_{2}. \ea

\indent The forms of $T$ and $W$ can be constructed with the
linear bases $T_{0},J_{0}$ and $W_{0}$. First we can write down
the most general possible structure of $W$, then the relations of
above $OPEs$ determine the coefficients of the terms in $W$, the
explicit results turn out to be very simple as follows:\\
\noindent(i)$\;\;s=3$
\ba
T &=T_0 ,\\
W &=W_{0}+\zeta_1 \partial^2 J_{0}+\zeta_2 \partial
J_{0}J_{0}+\zeta_3 J_{0}^{3}+\zeta_4 \partial T_{0}+\zeta_5
T_{0}J_{0}, \ea \noindent where
$$\align
\zeta_1&=6\zeta_0(2+3t^{2}+2t^{4}),\;\;\;\zeta_2=12 \sqrt{6}
\zeta_0 t(1+t^{2}),\\\allowdisplaybreak \zeta_3&=8\zeta_0
t^{2},\;\;\;\zeta_4=3\sqrt{6}\zeta_0
t(1+t^{2}),\;\;\;\zeta_5=12\zeta_0 t^{2},\\\allowdisplaybreak
\zeta_0&=\frac{1}{6t\sqrt{-15-34t^{2}-15t^{4}}}.
\endalign $$

\noindent (ii)$\;\;s=4$
\ba T =& T_0, \\
W =& W_0 + \eta_1 \partial^3 J_0 + \eta_2 \partial^2 J_0J_0 +
\eta_3(\partial J_0)^2 + \eta_4 \partial J_0(J_0)^2 +
\eta_5(J_0)^4\\
&+\eta_6 \partial ^2 T_0 + \eta_7(T_0)^2 + \eta_8 \partial T_0 J_0
+ \eta_9T_0 \partial J_0 + \eta_{10} T_0(J_0)^2, \ea \noindent
where
$$\align
\eta_1 =& \eta_0 (1800 + 5562 t^2 + 7744 t^4 + 6167 t^6 + 2631 t^8
+ 450 t^{10}),\\\allowdisplaybreak \eta_2 =& 12 \eta_0 t (450 +
1278 t^2 + 1429 t^4 +  752 t^6 + 150 t^8),\\\allowdisplaybreak
\eta_3 =& 6 \eta_0 t (1050 + 2932 t^2 + 3009 t^4 + 1353t^6 + 225
t^8),\\\allowdisplaybreak \eta_4 =& 12 \eta_0 t^2 (4 + 3 t^2)(150
+ 226 t^2 + 75 t^4),\\\allowdisplaybreak \eta_5 =& 6 \eta_0 t^3
(150 + 226 t^2 + 75 t^4),\\\allowdisplaybreak \eta_6 =& 3 \eta_0 t
(240 + 724 t^2 + 865 t^4 + 465 t^6 + 90 t^8),\\\allowdisplaybreak
\eta_7 =& 6 \eta_0 t^3 (3 + t^2)(32 + 27 t^2),\\\allowdisplaybreak
\eta_8 =& 12 \eta_0 t^2 (150 + 376 t^2 + 301 t^4 + 75
t^6),\\\allowdisplaybreak \eta_9 =& 12 \eta_0 t^2 (300 + 602 t^2 +
376 t^4 + 75 t^6),\\\allowdisplaybreak \eta_{10} =& 12 \eta_0 t^3
(150 + 226 t^2 + 75 t^4),\\\allowdisplaybreak \eta_0 =&
-(36t+12t^3)^{-1}(504000 + 3037560t^2 + 7617488 t^4 + 10300470 t^6
\\\allowdisplaybreak &+ 8109196 t^8 + 3716751 t^{10} + 918585
t^{12} + 94500 t^{14})^{-1/2}.
\endalign $$

\vskip 0.3in

{\bf\noindent  3. Ghost field realizations of the spinor non-critical $W_{2,s}$ strings} \\
\indent In this section, we construct the explicit ghost field
realizations of the spinor non-critical $W_{2,s}$ strings for
$s=3$ and $4$. Since the non-critical $W_{2,s}$ strings are the
theories of $W_{2,s}$ gravity coupled to a matter system, we
introduce the matter currents $T_{m}, W_{m}$ for the $W_{2,s}^m$
algebras in the beginning. Then we introduce the $(b,c)$ ghost
system with spins $(2,-1)$ for the spin-2 current, and the
$(\beta,\gamma)$ with spins $(s,1-s)$ for the spin-$s$ current.
The ghost fields $b,c,\beta,\gamma$ are all bosonic and commuting.
For the Liouville sector, the spinor field realizations of the
$W_{2,s}^L$ algebras were constructed in our work of Ref.[12]. We
can instead realize the $W_{2,s}^L$ algebras here by two pairs of
bosonic ghost fields $(b_{2},c_{2})$ and $(b_{3},c_{3})$. The
$BRST$ charge takes the form: \be
    Q_{B}=Q_{0}+Q_{1},
\ee \be
    Q_{0}=\oint dz\; c(T_{L}+T_{m}+KT_{bc}+yT_{\beta\gamma}),
\ee \be
    Q_{1}=\oint dz\; \gamma F(b_{2},c_{2},\beta,\gamma,T_{m},W_{m}),
\ee \noindent where $K,y$ are pending constants. The matter
currents $T_{m}$ and $W_{m}$, which have spin 2 and $s$
respectively, generate the $W_{2,s}^m$ algebras. The
energy-momentum tensors in Eq. (16) are given by \be
    T_{L}=2b_{2}\partial c_{2}+\partial b_{2}c_{2}+2b_{3} \partial c_{3}+ \partial b_{3}c_{3},
\ee
\be
    T_{bc}=2b\partial c+\partial bc.
\ee
\be
    T_{\beta\gamma}=s\beta\partial\gamma+(s-1)\partial\beta\gamma,
\ee \noindent The operator
$F(b_{2},c_{2},\beta,\gamma,T_{m},W_{m})$ has spin $s$ and ghost
number zero. The nilpotent $BRST$ charge generalizes the one for
scalar non-critical $W_{2,s}$ strings, and it is also graded with
$Q_{0}^{2}=Q_{1}^{2}=\{Q_{0},Q_{1}\}=0$. It should be emphasized
that the first condition is satisfied for any $s$ automatically,
we only need the other two conditions to determine $y$ and the
coefficients of the terms in
$F(b_{2},c_{2},\beta,\gamma,T_{m},W_{m})$.

\indent We now discuss the exact solutions of ghost field
realizations of the spinor non-critical $W_{2,3}$ and $W_{2,4}$
strings respectively, making use of the grading $BRST$ method and
the procedure mentioned above.\\

{\noindent  3.1. Ghost field realizations of the spinor non-critical $W_{2,3}$ strings} \\
\indent In this case, the $BRST$ operator $Q_{B}$ takes the form
of (15). Considering the most extensive combinations with correct
spin and ghost number, we can construct
$F(b_{2},c_{2},\beta,\gamma,T_{m},W_{m})$ in Eq. (17) as
following: \ba
                 F  &=f_{1}\beta^3\gamma^3 +f_{2} \partial\beta\beta\gamma^2
                     + f_{3} \partial \beta \partial \gamma
                     + f_{4} \beta \partial ^2 \gamma
                     + f_{5} b_{2}^3 c_{2}^3
                     + f_{6} \partial b_{2} b_{2} c_{2}^2
                     + f_{7} b_{2}^2 \partial c_{2} c_{2}
                     + f_{8} \partial b_{2} \partial c_{2}\\
                    &+ f_{9} \partial^2 b_{2} c_{2}
                     + f_{10} b_{2} \partial^2 c_{2}
                     + f_{11} \beta^2 \gamma^2 b_{2} c_{2}
                     + f_{12} \beta \gamma b_{2}^2 c_{2}^2
                     + f_{13} \partial \beta \gamma b_{2} c_{2}
                     + f_{14} \beta \partial \gamma b_{2} c_{2}\\
                    &+ f_{15} \beta \gamma \partial b_{2} c_{2}
                     + f_{16} \beta \gamma b_{2} \partial c_{2}
                     + f_{17} \beta \gamma T_{m}
                     + f_{18} b_{2} c_{2} T_{m}
                     + f_{19} \partial T_{m}
                     + f_{20} W_{m}. \\
\ea

\noindent Substituting this expression back into Eq. (17) and
using the nilpotency conditions, we can calculate the results of
$y$ and $f_{i}(i=1,2,...10)$. They correspond to three sets of solutions, i.e.\\
\noindent (i) $y=0$ and
$$\align
  f_{i}=&0 \;\;\;(i=5-12,17-20),\;\;\;  f_{13}=f_{16}, \;\;\; f_{14}=2f_{16}, \;\;\; f_{15}=f_{16},
\endalign $$
\noindent and $f_{j}(j=1-4,16)$ are arbitrary constants but do not
vanish at the same
time.\\
\noindent (ii) $y=1$ and
$$\align
           f_{1}=&\frac{1}{20}(-2f_{2} - f_{11}),\;\;\;f_{3} =\frac{1}{180}(480 f_{2} - 60 f_{4} - 90 f_{14} + 100 f_{15} + 80 f_{16}
                  - 3 C_m f_{18}-4 C_m f_{19}),\\\allowdisplaybreak
           f_{5} =&\frac{1}{72900}(27060 f_{2} - 39360 f_{4} - 8280 f_{8} + 15480 f_{9} + 67500 f_{11} -24480 f_{14}\\\allowdisplaybreak
                  &+ 46700 f_{15} + 2260 f_{16} - 591 C_m f_{18} - 1148 C_m f_{19}),\\\allowdisplaybreak
           f_{6}=&\frac{1}{6}(2 f_{8} -10 f_{9} + 5f_{14} - 10 f_{15}),\\\allowdisplaybreak
           f_{7}=&\frac{1}{2700}(5280 f_{2} - 7680 f_{4} - 2340 f_{8} - 1260 f_{9} + 1260 f_{14} + 1100 f_{15}\\\allowdisplaybreak
                  &- 3620 f_{16} - 33 C_m f_{18} - 224 C_m f_{19}),\\\allowdisplaybreak
           f_{10}=&\frac{1}{900}(5280 f_{2} - 7680 f_{4} - 540 f_{8} - 1260 f_{9} - 990 f_{14} + 1100 f_{15} + 880 f_{16} - 33 C_m f_{18} - 224 C_m f_{19}),\\\allowdisplaybreak
           f_{12}=&\frac{1}{1080}(-660 f_{2} + 960 f_{4} - 2700 f_{11} + 720 f_{14} - 1060 f_{15} - 380 f_{16} +  21 C_m f_{18} + 28 C_m f_{19}),\\\allowdisplaybreak
           f_{13}=&\frac{1}{540}(660f_{2}-960f_{4}-360f_{14}+700f_{15}+560f_{16}-21C_m f_{18}-28C_m f_{19}),\\\allowdisplaybreak
           f_{17}=&\frac{1}{10}(-3 f_{18} - 4 f_{19}),\\\allowdisplaybreak
\endalign $$
\noindent where $f_{j}(j=2,4,8,9,11,14,15,16,18,19,20)$ are
arbitrary
constants but do not vanish at the same time. $C_m$ is the central charge corresponding to the matter current $T_m$.\\
\noindent (iii) $y$ is an arbitrary constant and
$$\align
  f_{i}=&0 \;\;\;(i=5-12,17-20),\\\allowdisplaybreak
  f_{1}=&-\frac{2}{10}f_{2}, \;\;\;
  f_{3}=\frac{39}{16}f_{2}, \;\;\;
  f_{4}=\frac{11}{16}f_{2},\;\;\;
  f_{13}=f_{16}, \;\;\; f_{14}=2f_{16}, \;\;\; f_{15}=f_{16},\\\allowdisplaybreak
\endalign $$
\noindent where $f_{2}$ and $f_{16}$ are arbitrary
constants but do not vanish at the same time. \\
\indent Substituting the realizations (13) for $W_{2,3}$, with
$T_{0},J_{0}$ and $W_{0}$ given by (4,5,6), we obtain the
solutions for the ghost field realizations of the spinor non-critical $W_{2,3}$ string.\\

{\noindent  3.2. Ghost field realizations of the spinor non-critical $W_{2,4}$ strings} \\
\indent Similarly, for the case $s=4$, $Q_{B}$ also takes the form
of (15) and $F(b_{2},c_{2},\beta,\gamma,T_{m},W_{m})$ can be
expressed in the following form: \ba
              F & = g_{1}\beta ^4\gamma ^4+g_{2}\partial \beta \beta^2 \gamma ^3
                  +g_{3} \beta ^3 \partial \gamma \gamma ^2 + g_{4}\partial^2 \beta \beta \gamma ^2
                  +g_{5} \beta^2 \partial^2 \gamma \gamma + g_{6}(\partial \beta)^2\gamma ^2
                  +g_{7}\beta^2 (\partial \gamma)^2 \\
                & +g_{8} \partial \beta \beta \partial\gamma \gamma
                  +g_{9}\partial^3 \beta \gamma + g_{10}\beta \partial^3 \gamma
                  +g_{11} b_{2}^4 c_{2}^4 + g_{12}\partial b_{2} b_{2}^2 c_{2}^3
                  +g_{13} b_{2}^3 \partial c_{2} c_{2}^2  +g_{14} \partial^2 b_{2} b_{2} c_{2}^2\\
                & +g_{15}  b_{2}^2 \partial^2 c_{2} c_{2}+g_{16} (\partial b_{2})^2  c_{2}^2
                  +g_{17} b_{2}^2  (\partial c_{2})^2
                  +g_{18} \partial b_{2} b_{2} \partial c_{2} c_{2}+g_{19} \partial^3 b_{2} c_{2}
                  +g_{20} b_{2} \partial^3 c_{2}\\
                & +g_{21}\beta^3 \gamma^3 b_{2} c_{2}
                  +g_{22} \beta^2 \gamma^2 b_{2}^2 c_{2}^2
                  +g_{23} \beta \gamma b_{2}^3 c_{2}^3+g_{24}\partial \beta \beta \gamma^2 b_{2} c_{2}
                  +g_{25} \beta ^2 \partial \gamma \gamma b_{2} c_{2}
                  +g_{26} \beta ^2 \gamma^2 \partial b_{2} c_{2}\\
                & +g_{27} \beta ^2 \gamma^2 b_{2} \partial c_{2}
                  +g_{28} \partial \beta \gamma b_{2}^2 c_{2}^2+g_{29} \beta \partial \gamma b_{2}^2 c_{2}^2
                  +g_{30} \beta \gamma \partial b_{2} b_{2} c_{2}^2+g_{31} \beta \gamma b_{2}^2 \partial c_{2}c_{2}
                  +g_{32} \partial\beta \partial\gamma b_{2} c_{2}\\
                & +g_{33} \beta \gamma \partial b_{2} \partial c_{2}+g_{34} \partial\beta \gamma \partial b_{2} c_{2}
                  +g_{35} \partial \beta \gamma b_{2} \partial c_{2}+g_{36} \beta \partial \gamma \partial b_{2} c_{2}
                  +g_{37} \beta \partial \gamma b_{2} \partial c_{2}+g_{38} \partial^2 \beta \gamma b_{2} c_{2}\\
                & +g_{39} \beta \partial^2\gamma b_{2} c_{2}
                  +g_{40} \beta \gamma \partial^2 b_{2} c_{2}+g_{41} \beta \gamma b_{2} \partial^2 c_{2}
                  + g_{42} \beta^2 \gamma^2 T_{m}
                  + g_{43} \partial \beta \gamma T_{m}+ g_{44} \beta \partial \gamma T_{m}\\
                & + g_{45} \beta \gamma \partial T_{m} + g_{46} b_{2}^2 c_{2}^2 T_{m}
                  + g_{47} \partial b_{2} c_{2} T_{m}
                  + g_{48} b_{2} \partial c_{2} T_{m}
                  + g_{49} b_{2} c_{2} \partial T_{m}\\
                & + g_{50} \beta \gamma b_{2} c_{2} T_{m}
                  + g_{51}T_{m}^2 + g_{52}\partial^2 T_{m} +g_{53}W_{m} . \\
\ea \indent There are three sets of solutions:\\
\noindent (i) $y=0$ and
$$\align
        g_{i} &=0\;\;\; (i=11-23,39,42,46-53),\\ \allowdisplaybreak
        g_{24} &= 2 g_{26},\;\;\; g_{25} = 3 g_{26},\;\;\; g_{27} = g_{26}, \;\;\;g_{28} =\frac{1}{2} g_{30},\;\;\; g_{29} = g_{30},\;\;\;g_{31} = g_{30}, \\\allowdisplaybreak
        g_{32} &= 2 (g_{34} -g_{40}),\;\;\; g_{35} = g_{33} + g_{34} - 2 g_{40}, \;\;\;g_{36} = 2 g_{40},\;\;\;g_{37} = 2 (g_{33} - g_{40}),\\\allowdisplaybreak
        g_{38} &= g_{34} - g_{40},\;\;\; g_{41} = g_{33} - g_{40},\;\;\;g_{43} = g_{45},\;\;\; g_{44} = 2 g_{45},\\\allowdisplaybreak
\endalign $$
\noindent where $g_{j} (j=1-10,26,30,33,34,40,45)$ are arbitrary
constants but do not vanish at the same
time.\\
\noindent (ii) $y=1$ and
$$\align
  g_{1} &= \frac{1}{140} (-8 g_{2} + 6 g_{3} - 3 g_{21}),\;\;\;g_{10} = \frac{1}{12}(28 g_{5} - 14 g_{7} + 3 g_{39}),\;\;\;
  g_{9} = \frac{1}{96} (-90 g_{4} - 108 g_{5}  \\\allowdisplaybreak
        & + 24 g_{7} + 30 g_{8} - 6 g_{33} + 10 g_{36} + 8 g_{37} - 6 g_{39} - 14 g_{40} -  10 g_{41} + C_m g_{44} - 2 C_m g_{45}),\\ \allowdisplaybreak
  g_{12} &= \frac{1}{9720}(-258720 g_{2} + 194040 g_{3} - 20610 g_{4} + 99364 g_{5} - 108768 g_{6} - 92808 g_{7} + 43126 g_{8}- 8424 g_{14}\\ \allowdisplaybreak
        &  + 23520 g_{26} - 23520 g_{27} - 3558 g_{33} + 6592 g_{34}  - 6592 g_{35}- 13662 g_{36}- 1620 g_{37} + 918 g_{39}\\ \allowdisplaybreak
        &  + 24290 g_{40} + 13390 g_{41}   - 2940 C_m g_{42}- 824 C_m g_{43} - 927 C_m g_{44} + 2678 C_m g_{45} +  108 C_m g_{47}),\\ \allowdisplaybreak
  g_{15} &= \frac{7}{6} (g_{37} - g_{39} - 2 g_{41}),\;\;\;
  g_{18} = \frac{7}{3} (-2 g_{33} + g_{36} + g_{37} - 2 g_{39}),\\ \allowdisplaybreak
  g_{19} &= \frac{1}{18} (-6 g_{14} + 7 g_{36} - 7 g_{39} - 14 g_{40}),\\ \allowdisplaybreak
  g_{24} &= \frac{1}{9} (-80 g_{2} + 60 g_{3} - 6 g_{4} - 36 g_{5} - 48 g_{6} + 18 g_{8} + 10 g_{26} + 8 g_{27} + C_m g_{42}),\\ \allowdisplaybreak
  g_{25} &= \frac{1}{6} (88 g_{2} - 66 g_{3} + 48 g_{4} - 16 g_{5} + 24 g_{7} - 16 g_{8} + 10 g_{26} + 8 g_{27} + C_m g_{42}),\\ \allowdisplaybreak
  g_{31} &= \frac{1}{3} (21 g_{26} - 21 g_{27} + 3 g_{30} - 3 g_{33} + 4 g_{34} - 4 g_{35} - 3 g_{36} + 3 g_{37} + 5 g_{40} + g_{41}),\\ \allowdisplaybreak
  g_{46} &= \frac{1}{54} (882 g_{42} + 58 g_{43} - 113 g_{44} + 168 g_{45} -43 g_{47} - 11 g_{48} - 22 g_{51} - 5 C_m g_{51}),\\ \allowdisplaybreak
  g_{49} &= \frac{1}{18} (110 g_{43} - 13 g_{44} - 84 g_{45} + 25 g_{47} +  20 g_{48} + 22 g_{51} + 5 C_m g_{51}),\\ \allowdisplaybreak
  g_{50} &= \frac{2}{3} (-21 g_{42} - 4 g_{43} + 3 g_{44} - 2 g_{45}),\;\;\;g_{52} = \frac{1}{12} (-22 g_{43} + 11 g_{44} - 5 g_{47} - 4 g_{48} -  8 g_{51} - C_m g_{51}),\\ \allowdisplaybreak
\endalign $$
\noindent where $g_{j}
(j=2-8,14,21,26,27,30,33-37,39-45,47,48,51,53)$ are arbitrary
constants but do not vanish at the same time. The other
coefficients $g_{i}(i=12,13,16,17,20,22,23,28,29,32,38)$ have a
more verbose form which is similar to $g_{12}$, we don't
list them here.\\
\noindent (iii) $y$ is an arbitrary constant and
$$\align
        g_{i} &=0\;\;\; (i=11-23,39,42,46-53),\\ \allowdisplaybreak
        g_{1} &= \frac{289}{22330}(3 g_{4} + 2 g_{5} - g_{8}),\;\;\;
        g_{2} = \frac{1}{1276}(957 g_{3} - 289 (3 g_{4} + 2 g_{5} - g_{8})),\\ \allowdisplaybreak
        g_{6} &= \frac{1}{7656}(7713 g_{4} + 38 g_{5} - 19 g_{8}),\;\;\;
        g_{7} = \frac{1}{116}(57 g_{4} + 270 g_{5} - 19 g_{8}),\\ \allowdisplaybreak
        g_{9} &= \frac{63}{232} (-3 g_{4} - 2 g_{5} + g_{8}),\;\;\;
        g_{10} = \frac{133}{696} (-3 g_{4} - 2 g_{5} + g_{8}),\\ \allowdisplaybreak
        g_{24} &= 2 g_{26},\;\;\; g_{25} = 3 g_{26},\;\;\; g_{27} = g_{26}, \;\;\;g_{28} =\frac{1}{2} g_{30},\;\;\; g_{29} = g_{30}, \\\allowdisplaybreak
        g_{31} &= g_{30},\;\;\;g_{32} = 2 (g_{34} -g_{40}),\;\;\; g_{35} = g_{33} + g_{34} - 2 g_{40}, \\\allowdisplaybreak
        g_{36} &= 2 g_{40},\;\;\;g_{37} = 2 (g_{33} - g_{40}),\;\;\; g_{38} = g_{34} - g_{40},  \\\allowdisplaybreak
        g_{41} &= g_{33} - g_{40},\;\;\;g_{43} = g_{45},\;\;\; g_{44} = 2 g_{45},\\\allowdisplaybreak
\endalign $$
\noindent where $g_{j} (j=3,4,5,8,26,30,33,34,40,45)$  are
arbitrary constants but do not vanish at the same
time.\\
\indent Substituting the realizations (14) for $W_{2,4}$, with
$T_{0},J_{0}$ and $W_{0}$ given by (4,5,7), we obtain the
solutions for the ghost field realizations of the spinor
non-critical
$W_{2,4}$ string.\\
\indent Since the solutions of
$F(b_{2},c_{2},\beta,\gamma,T_{m},W_{m})$ are of general type, the
constructions of $Q_1$ are general naturally. Noting that the
constructions of the energy-momentum tensors in $Q_0$ include all
combinations of various fields, the realizations of $Q_B=Q_0+Q_1$
that we have obtained are of the most general constructions. For
the case of the spinor critical $W_{2,s}$ strings, the
realizations can be obtained when the matter currents $T_{m}$ and
$W_{m}$ in Eqs. (16, 17) vanish. It is easy to see that the
constructions become relatively simple and there are also three
sets of solutions for each $s$. For the reason of space, We don't list them here.\\
\indent Note that our new realizations, based on spinor $W_{2,s}$
strings, are very different from that of the scalar $W_{2,s}$
strings. The critical center charges $C$ for scalar $W_{2,s}$ are
fixed, for example $C=100$ for $s=3$ and $C=176$ for $s=4$. One
usually need an exact center charge for the Liouville sector to
obtain the realizations of certain scalar non-critical $W_{2,s}$
string. But for spinor $W_{2,s}$ strings, the center charges are
arbitrary for both the critical and non-critical cases. The reason
is that the condition $Q_{0}^2=0$ is satisfied automatically as a
result of the property of spinor field.\\

{\bf\noindent 4. Conclusion}\\
\indent In conclusion, we have constructed the new ghost field
realizations of the $W_{2,3}$ and $W_{2,4}$ algebras, making use
of the fact that the $W_{2,3}$ and $W_{2,4}$ algebras can be
linearized through the addition of a spin-1 current. The spin-s
current $W_{0}$ in the linearized $W_{1,2,s}$ algebras is null.
This null current is identically zero in the case of the
two-spinor realizations of the $W_{1,2,s}$ algebras [16]. However,
$W_{0}$ can be non-zero and is realized by the bosonic ghost
fields in this paper. In addition, the other two currents $T_{0}$
and $J_{0}$ are also realized by the bosonic ghost fields. Then
the non-linear bases of $W_{2,s}$ have been constructed with these
linear bases $T_0, J_0$ and $W_0$. Subsequently, we use these new
realizations to build the nilpotent $BRST$ charges of spinor
non-critical $W_{2,s}$ strings. The $BRST$ charge generalizes the
one for scalar non-critical $W_{2,s}$ strings, and it is graded
with $Q_{0}^{2}=Q_{1}^{2}=\{Q_{0},Q_{1}\}=0$. In particular, using
the procedure mentioned above, we have already discussed the cases
of $s=3$ and $4$ in detail. It is easy to see that these
constructions are very standard, that is, there are three
solutions for $s=3$ and $4$, respectively. It is worth to point
out that the results of the non-critical $W_{2,s} (s=3,4)$ strings
would turn to that of the corresponding critical strings when the
matter currents vanish. Similarly, other field realizations of
$W_{2,s}$ strings can be obtained with our method. And
furthermore, such realizations with higher spin $s$ will be
expected to exist. In view of these points, it would be
interesting to investigate the physical states, which are defined
as the cohomology classes of a nilpotent $BRST$ operator $Q_B$.
Moreover other implications, such as the construction of the
W-gravity, can be studied also.\\

{\bf \noindent Acknowledgements}

\indent It is a pleasure to thank Prof. Y.S. Duan for useful
discussions. We have also made extensive use of a Mathematica
package for calculating $OPEs$, written by Prof. K. Thielemans
[17].

\end{document}